\documentclass[prl, twocolumn, showpacs, preprintnumbers,amsmath,amssymb,superscriptaddress]{revtex4-1}
\usepackage{graphicx}
\usepackage{dcolumn}
\usepackage{bm}
\usepackage{subfigure}
\usepackage{multirow}
\usepackage{amsmath}
\usepackage{color}
\usepackage[font=footnotesize, justification=raggedright,singlelinecheck=false]{caption}
\begin{document}

\title{Role of local assembly in the hierarchical crystallization of associating colloidal hard hemispheres}

\author{Qun-li Lei}
 \affiliation{%
 School of Chemical and Biomedical Engineering, Nanyang Technological University, 637459, Singapore
 }%
 
\author{Kunn Hadinoto}
 \affiliation{%
 School of Chemical and Biomedical Engineering, Nanyang Technological University, 637459, Singapore
 }%

\author{Ran Ni}
 \email{r.ni@ntu.edu.sg}
 \affiliation{%
 School of Chemical and Biomedical Engineering, Nanyang Technological University, 637459, Singapore
 }%

\begin{abstract}
Hierarchical self-assembly consisting of local associations of simple building-blocks for the formation of complex structures widely exists in nature, while the essential role of local assembly remains unknown. 
In this work, by using computer simulations, we study a simple model system consisting of associating colloidal hemispheres crystallizing into face-centered-cubic crystals comprised of spherical dimers of hemispheres, focusing on the effect of dimer formation on the hierarchical crystallization.
We found that besides assisting the crystal nucleation because of increasing the symmetry of building-blocks, the association between hemispheres can also induce both re-entrant melting and re-entrant crystallization depending on the range of interaction. Especially when the interaction is highly sticky, we observe a novel re-entrant crystallization of identical crystals, which melt only in certain temperature range. This offers a new axis in fabricating responsive crystalline materials by tuning the fluctuation of local association. 
\end{abstract}


\maketitle

Hierarchical self-assembly, where the products from the lower-level assembly  act as building-blocks for the higher-level self-assembly, is first used by nature to accurately build complex micro-structures~\cite{elemans2003mastering,pieters2016natural,Alberts2002}. The processes are usually accompanied with the formation of local assemblies, e.g. dimerization~\cite{marian2004power,matthews2012p}, with which higher level complex structures can be built with ease~\cite{ahnert2015p,king2013practical,goodsell2000s,zhang2014design,bale2016accurate}. For example, in the self-assembly of icosahedral virus capsids, anisotropic protein monomers first form dimers to gain centrosymmetry, then the dimers assemble into pentamer blocks, which crystallize into  ``spherical crystals''~\cite{krishnamani2016,baschek2012virus}. 
Accordingly, a new  \emph{racemic protein crystallography}~\cite{yeates2012racemic}  method was recently proposed, where synthesized enantiomers or enontiomorphs are used to co-crystallize some natural chiral proteins, whose crystals are difficult to obtain using traditional crystallography.~\cite{laganowsky2011a}.

In colloidal self-assembly, one of the major tasks is to design anisotropic particles to fabricate crystalline materials with desired properties~\cite{glotzer2007,hanrev2016,glotzer2012,escobedo2011}. It was recently suggested that for self-assembly of complex colloidal crystals, one can pre-assemble the local structures to help the hierarchical crystallization~\cite{hynninen2007self,dijkstra2016,ducrot2017c}. However, the role of the local assembly for the hierarchical crystallization remains unclear. Here we investigate the hierarchical crystallization of a simple yet representative system consisting of associating colloidal hemispheres without centrosymmetry, which at high density self-assemble into a face-centered-cubic (FCC) crystal of spherical dimers of hemispheres, i.e. FCC$^2$ crystal. We found that besides assisting the hierarchical nucleation of FCC$^2$ crystal of colloidal hemispheres, the formation of local assemblies can induce, depending on the interaction range of association, both re-entrant melting and re-entrant crystallization of FCC$^2$ crystals within certain density range. This suggests a new way of fabricating responsive photonic materials by controlling local structural fluctuations.

\begin{figure}[b]
\centering
		\resizebox{80mm}{!}{\includegraphics[trim=0.0in 0.0in 0.0in 0.0in]{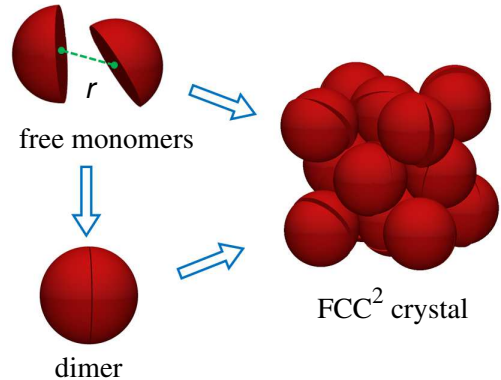} } 
\caption{\label{fig1} (Color online) Schematic illustration of the model: free colloidal hemisphere monomers can self-assemble into an FCC crystal consisting of spherical dimers. }
\end{figure}

\begin{figure*}[t]
\centering
		\resizebox{150mm}{!}{\includegraphics[trim=0.5in 0.0in 0.0in 0.0in]{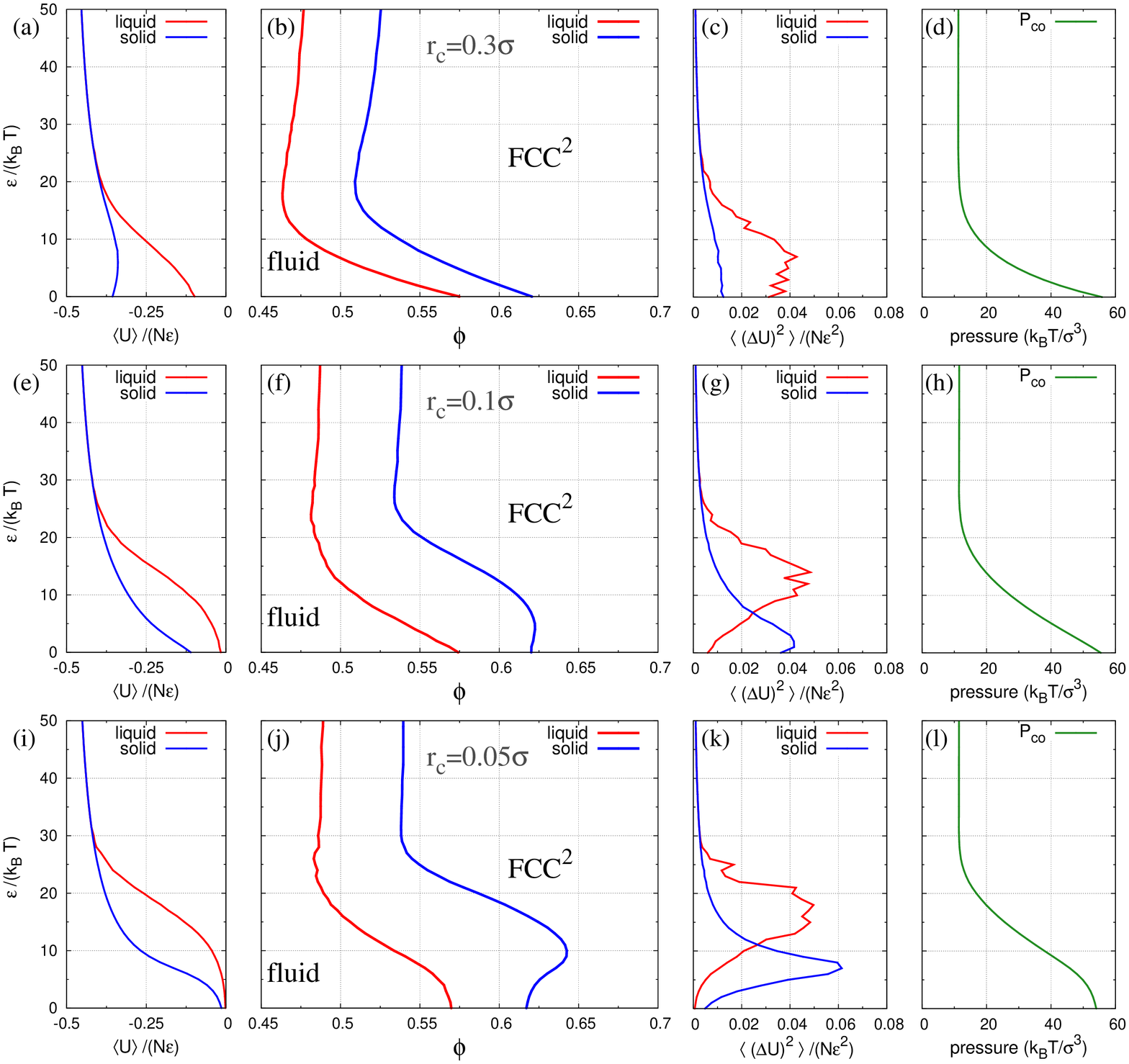} } 
\caption{(Color online) Phase diagrams of associating colloidal hemispheres for various interaction ranges, i.e. $r_c = 0.3\sigma$ (b), $0.1\sigma$ (f) and $0.05\sigma$ (j), in the representation of volume fraction vs. inverse temperature $\epsilon/k_B T$. 
The average energy per particle ${\langle U \rangle}/N{\epsilon}$, normalized energy fluctuation ${\langle \Delta U \rangle^2 }/{N\epsilon^2}$, and the co-existing pressure $P_{co}$ are shown in column one (a,e,i), three (c,g,k) and four (d,h,l), respectively.
\label{fig2}}
\end{figure*}

We consider a system of $N$ colloidal hard hemispheres, which at high density can crystallize into an FCC$^2$ crystal~\cite{marechal2010p1,marechal2010p2,mcbride2017p}. To control the formation of local structures, i.e. spherical dimers, we introduce an attraction between hemispheres. The total energy of the system is given by
\begin{equation}
	U = \sum_{i<j} U_{HHS}(i,j) + U_b(i,j).
\end{equation}
$U_{HHS}(i,j)$ is the hard-core potential between hemisphere $i$ and $j$ with $U_b(i,j)$ the attraction given by
\begin{eqnarray}
U_{b}(i,j) =\left\{
\begin{array}{rl}
& \epsilon \left(\frac{r_{ij}}{r_c}-1\right) \quad \quad \quad \quad { (r_{ij}  \leq r_{c}) }\\ 
&0 \quad \quad \quad \quad \quad \quad \ \  \ \  \ \ { (r_{ij}>r_{c}) },\\
\end{array}
\right.
\end{eqnarray}
where $r_{ij}$ is the center-to-center distance between the flat surfaces of hemisphere $i$ and $j$ (Fig.~\ref{fig1}). 
To ensure the attraction only exist between the flat surfaces of two hemispheres, here we choose $r_c \le 0.3 \sigma$ with $\sigma$ the diameter of hemisphere. The reduced temperature $T^*={k_BT}/{\epsilon}$ controls the associating degree, or the  dimer fraction $\theta$, with $k_B$ and $T$ the Boltzmann constant and temperature of the system, respectively. Here a spherical dimer is defined as a collection of two hemispheres, whose center-to-center distance is smaller than $r_c$. In the limit of $T^* \rightarrow 0$, all hemispheres form spherical dimers in the fluid  recovering a system of hard spheres.~\cite{hard_sphere}. 

\begin{figure*}[t]  
\centering
		\resizebox{150mm}{!}{\includegraphics[trim=0.5in 0.0in 0.0in 0.0in]{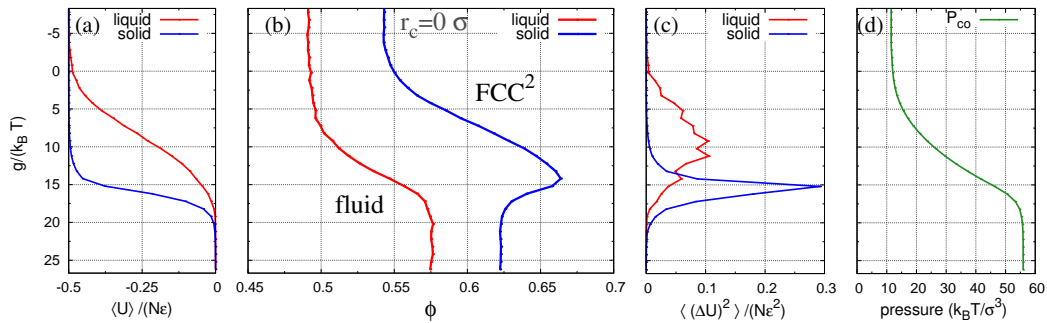} } 
\caption{
(Color online) Phase diagrams of sticky colloidal hemispheres, i.e. $r_c = 0$ (b), in the representation of volume fraction vs. association strength $g/k_B T$. Corresponding average energy per particle ${\langle U \rangle}/{N\epsilon}$, normalized energy fluctuation ${\langle \Delta U \rangle^2 }/{N\epsilon^2}$, and the co-existing pressure $P_{co}$ are shown in (a), (c) and (d), respectively. \label{fig3}}
\end{figure*}

We first calculate the phase diagram for the system of hard hemispheres, i.e. $\epsilon/k_B T = 0$, by using the Einstein integration, where all particles are modelled as penetrable repulsive hemispheres and each particle is attached to a crystalline lattice site via a spring. By increasing the strength of the spring and decreasing the strength of repulsion, the system recovers a non-interacting Einstein plastic crystal~\cite{plastic_crystal}. However, different from conventional plastic crystals, in the FCC$^2$ crystal of hemispheres, the two particles on the same lattice site are exchangeable contributing a free energy of $\frac{1}{2}\ln{2} k_B T$ per particle, and the resulting free energy of the FCC$^2$ Einstein crystal is 
\begin{equation}\label{eq3}
\frac{F_{Einst}}{k_B T} = -\frac{3(N-1)}{2} \ln \left( \frac{\pi k_B T}{\lambda_{max}} \right) + \ln \left( \frac{\sigma^3}{V N^{1/2}} \right) + \frac{N}{2} \ln 2,
\end{equation}
where $\lambda_{max}$ is the strength of spring with $V$ the volume of the system. This extra free energy contribution from indistinguishness generally exists in all hierarchical plastic crystals, whose building-blocks are local assemblies of smaller particles. By using this Einstein crystal combined with thermodynamic integrations~\cite{bookfrenkel}, we obtain $[\phi_{f},\phi_{FCC^2}] = [0.574,0.622]$ with $\phi_{f}$ and $\phi_{FCC^2}$the co-existing packing fraction of the fluid and FCC$^2$ crystal, respectively. These are substantially higher than the values obtained in Ref.~\cite{mcbride2017p}, and the reason is the last term of Eq.~\ref{eq3} missing in the previous works~\cite{sup_info}, as by reducing the free energy of our crystal phase by  $\frac{1}{2}\ln{2} k_B T$ per particle, we obtain the same phase boundaries as in Ref.~\cite{mcbride2017p}. 

Next we trace the change of phase boundaries as a function of $\epsilon/k_BT$ by using the Gibbs-Duhem integration 
\begin{equation}\label{gdi}
\left[ \frac{ \mathrm{d} \ln P}{\mathrm{d} (\epsilon/k_B T)} \right]_{coex} = - \frac{\Delta h }{ P \Delta v \epsilon/k_B T},
\end{equation}
where $\Delta h$ and $\Delta v$ are the difference of enthalpy and specific volume between two coexisting phases, respectively. We perform isobaric-isothermal Monte Carlo simulations with $N = 1,000$ hemispheres to solve Eq.~\ref{gdi} starting from the system of hard hemispheres, i.e. $\epsilon/k_B T = 0$, and the resulting phase diagrams for various attraction ranges are shown in the second column of Fig.~\ref{fig2}. For the case of relatively long range attraction, i.e., $r_c=0.3\sigma$, one can see that with increasing $\epsilon/k_BT$ from 0, both the phase boundaries of fluid and FCC$^2$ phases first decrease and then increase approaching the limit of hard-sphere systems. They reach $[\phi_{f},\phi_{FCC^2}] = [0.47, 0.52]$ at an intermediate association $\epsilon/k_B T \simeq 17$, which are even lower than those of hard-sphere systems. This non-monotonic behaviour of the crystallization packing fraction implies an interesting re-entrant melting at certain packing fraction range with increasing the attraction. Moreover, from the hard-sphere limit, with decreasing the attraction, the phase boundaries shift to lower values, which suggests that at fixed packing faction, the crystal nucleation rate in the fluid increases when the spherical dimers have certain shape fluctuations. 
With decreasing the attraction range $r_c$, the re-entrant melting becomes weaker, and it almost disappears at $r_c = 0.1$ and $0.05\sigma$. Surprisingly, when the $r_c$ is very small, i.e. $0.05 \sigma$, the melting packing fraction of the FCC$^2$ crystal changes non-monotonically when approaching the system of hard hemispheres, and it reaches the maximal value of $\phi_{FCC^2} \simeq 0.64$ at $\epsilon/k_BT = 10$. With further decreasing the attraction, the melting line of FCC$^2$ crystal moves down to $\phi_{FCC^2} \simeq 0.62$ at the hard-hemisphere limit. This non-monotonic behaviour of $\phi_{FCC^2}$ suggests that at certain fixed packing fraction between 0.62 and 0.64, by increasing the strength of short range attraction, the system undergoes a novel re-entrant crystallization by forming  identical crystals at both strong and weak attraction limits which melt at certain intermediate attraction. However, although the re-entrant melting and re-entrant crystallization both exist in the system of associating colloidal hemispheres depending on the associating range, the co-existing pressure always monotonically decreases with increasing $\epsilon/k_B T$ (Fig.~\ref{fig2} right column). Additionally, by using the Gibbs-Duhem integration from hard-hemisphere systems with increasing attraction, we reproduce the phase boundary of hard sphere systems at $\epsilon/k_B T \rightarrow \infty$, which verifies our free energy calculation of hard hemisphere systems.
Here we focus on the phase transition between fluid and the FCC$^2$ crystal, and full phase diagrams can be found in Ref~\cite{sup_info}.

To understand the physics behind these re-entrant behaviours, we plot average energy per particle $\langle U \rangle / N \epsilon $ and the energy fluctuation $\langle \Delta U^2 \rangle /N\epsilon^2 $ on the fluid-FCC$^2$ phase boundaries jointly with the phase digram in the first and third column of Fig.~\ref{fig2}. 
In the systems of short range attractive hard hemispheres, the change of $\langle U \rangle / N \epsilon $ is very similar to that of $\theta$~\cite{sup_info}.
As shown in Fig.~\ref{fig2}a, e, and i, for $r_c = 0.05, 0.1$ and $0.3\sigma$, $\langle U \rangle /N \epsilon $ of co-existing phases matches with each other at high attraction strength, where all hemispheres form dimers. Decreasing $\epsilon/k_B T$ increases $\langle U \rangle /N \epsilon $ of co-existing phases gradually, which implies that the average distance between two hemispheres in spherical dimers increases. This change has little influence on the phase boundary when $r_c$ is small, i.e. $0.05\sigma$. However, in the system of relatively longer range attraction, i.e. $r_c=0.3\sigma$, this effectively increases the ``size'' of the spheres moving the phase boundary to lower packing fractions. Further decreasing the attraction induces deviation between $\langle U \rangle /N \epsilon $ in the two co-existing phases, and the energy of fluid increases faster than solid indicating that the dissociation of spherical dimers occurs first in the fluid. 
This imbalance implies that the fluid gains more entropy from the dissociation than solid. Then the co-existing packing fractions shift to high values to equalize the chemical potentials of co-existing phases. This effect, along with the increased number of free hemispheres, explains the increase of co-existing pressure as well as the re-entrant melting. Moreover, as shown in Fig.~\ref{fig2}c, g, and k, the energy fluctuations on the coexisting phases, especially in the coexisting FCC$^2$ crystal, changes differently with decreasing attraction for different $r_c$. When the attraction range is relatively long, i.e. $r_c = 0.3\sigma$, the energy fluctuation $\langle \Delta U^2 \rangle /N\epsilon^2 $ increases monotonically when decreasing the attraction strength, while at short range attractions, it develops a maxima when approaching the hard hemisphere limit. Interestingly, the location of the energy fluctuation maxima is very close to the maximal melting packing fraction of FCC$^2$ crystals leading to the re-entrant crystallization of identical FCC$^2$ crystals with increasing attraction.

\begin{figure}[t]
\centering
		\resizebox{80mm}{!}{\includegraphics[trim=0.0in 0.0in 0.0in 0.0in]{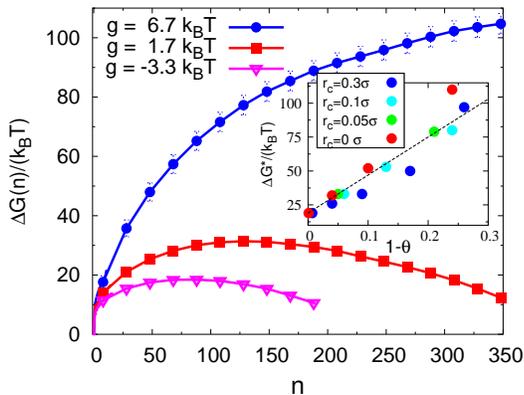} } 
\caption{(Color online) Nucleation barrier of FCC$^2$ crystals $\Delta G(n)/k_B T$ as a function of nucleus size $n$ in systems of sticky colloidal hemispheres, i.e. $r_c = 0$, with various association strength $g$ at the supersatruation of $0.54k_BT$ per spherical dimer. Inset: the heights of nucleation barrier as a function of the free monomer fraction $1-\theta$ for various attraction strength and interaction range at the supersatruation of $0.54k_BT$ per spherical dimer, where the dash line is to guide the eye.\label{fig4}}
\end{figure}

To further explore the nature of this intriguing re-entrant crystallization, we simulate a system of associating hard hemispheres with $r_c \rightarrow 0$. In this limit, to bind two hemispheres forming a spherical dimer, $\epsilon/k_B T$ needs to approach infinity, and the dimerization fraction $\theta = - 2\langle U \rangle / N \epsilon $. Therefore, instead of $\epsilon/k_B T$,we define a dimerization free energy $g$ to describe the association strength between hemispheres as
\begin{equation}
g = -k_B T \ln{Z_b} = - k_B T\ln\left\{ \int \exp \left[-\beta U_b(r) \right] d\mathbf{s}  \right\},
\end{equation} 
where $Z_b$ can be seen as the internal partition function of a spherical dimer with $\mathbf{s}$ the internal degrees of freedom of two hemispheres. Since the entropic barrier for dimerization increasing dramatically when $r_c \rightarrow 0$, we devise a modified aggregation-volume-bias Monte Carlo algorithm ~\cite{avbmc} to accelerate the simulation ~\cite{sup_info}, and the results are shown in Fig.~\ref{fig3}. Compared with $r_c = 0.05 \sigma$, a more pronounced re-entrant crystallization is observed in systems with $r_c \rightarrow 0$ accompanied with the larger energy difference between the two co-existing phases suggesting a large difference of the dimer fraction in the two phases.
Especially, as shown in Fig.~\ref{fig3}a, when the dimer fraction decreases to 25\% in fluid, all particles in the FCC$^2$ crystal still remain dimerized. A small further increase of temperature induces a sharp change of energy in the FCC$^2$ crystal, and a pronounced energy fluctuation peak appears suggesting a collective dissociation in the crystal, which is stronger at smaller $r_c$.
This collective behaviour can be seen as a kind of weak solid-solid transition from high density to low density similar to the solid-solid transition in systems of sticky hard spheres~\cite{bolhuis1994iso}. However, in our systems of sticky hard hemispheres, the nature of dissociation of spherical dimers is continuous, and not strong enough to drive a first order phase separation, but produces a new re-entrant crystallization in the system to form identical crystals with changing temperature. 

Furthermore, we study the nucleation of FCC$^2$ crystals from the fluids of colloidal hemispheres. We perform umbrella sampling Monte Carlo simulations~\cite{filion2010,ran2010jcp} to calculate the free energy barrier $\Delta G(n) / k_B T = -\ln P(n)$ with $P(n)$ the probability of finding a nucleus containing $n$ solid crystal-like dimers, which is determined by using the bond orientation order parameter~\cite{bop,sup_info}. The obtained nucleation barriers for systems at the supersaturation of $|\Delta \mu| = |\mu_{FCC^2} - \mu_{fluid}| = 0.54 k_BT$ per spherical dimer with various association strength at $r_c = 0$ are shown in Fig.~\ref{fig4}. One can see that with decreasing the association strength $g/k_BT$, at the same supersaturation, the nucleation barrier dramatically increases. As shown in the inset of Fig.~\ref{fig4}, nucleation barrier heights of systems with different interaction ranges change very similarly with decreasing the fraction of spherical dimers in the supersaturated fluids. 
This suggests that the determining factor for the nucleation rate of FCC$^2$ crystal is the fraction of spherical dimers in the fluid, while the exact form of interaction is less important. Moreover, as our simulations are performed at the constant supersaturation, the higher nucleation rate in stronger two-step hierarchical self-assembling systems cannot be explained by the increase of driving force.
Instead, our results demonstrate that for particles of low-symmetry, like hemispheres, locally self-assembling into secondary building-blocks of high-symmetry can dramatically increase the self-assembly efficiency~\cite{glotzer1,glotzer2}.
This gives a generic explanation on why dimerization or local structural formation is usually the first step in the protein self-assembly, and why racemic protein crystallography works better by introducing local association of enantiomers~\cite{yeates2012racemic}.

In conclusion, by performing computer simulations for a simple yet representative model system of colloidal hemispheres, we investigate the role of local assembly in hierarchical crystallization. 
We found that depending on the range of attraction driving the formation of local structures, i.e. spherical dimers, the system posses novel re-entrant melting and re-entrant crystallization at certain densities. Especially in the system of the sticky colloidal hemispheres, i.e. $r_c \rightarrow 0$, where the exact form of attraction is not important, increasing the strength of attraction can induce a new re-entrant crystallization by forming identical FCC$^2$ crystals at both weak and strong attraction limits which melts at intermediate attraction strength. This is due to the collective dissociation of  spherical dimers. 
We argue that this  sticky association induced new re-entrant crystallization generally should exist in many hierarchical self-assembling systems, and more subunits in each local assembly can produce stronger re-entrant crystallization, which could be interesting for future investigations.
In experiments, such sticky attraction, for example, can be realized by using hydrophobic coatings on the flat surface of colloidal hemispheres~\cite{granick1,granick2,granick3}, which may open up a new way of making novel responsive photonic materials~\cite{resp_PC}. 
Moreover, we also studied the nucleation of FCC$^2$ crystal from supersaturated fluids, and we demonstrated that at the same supersaturation, the increase of the fraction of spherical dimers in fluids significantly lowers the nucleation barrier suggesting that the existence of pre-assembled local structures is of primary importance for the hierarchical crystallization, which is relevant for designing the self-assembly of anisotropic colloids~\cite{ducrot2017c} and protein crystallization~\cite{yeates2012racemic}. Our results lay the first stone in understanding the role of local structural formation in the multi-scale hierarchical assembly, and a number of interesting questions can be further explored in this direction, e.g. the effect of local structural fluctuations on hierarchical glass transitions~\citep{speckprl2012}.

\begin{acknowledgments}
This work is supported by Nanyang Technological University Start-Up Grant (NTU-SUG: M4081781.120), Academic Research Fund Tier 1 from Singapore Ministry of Education (M4011616.120), the  Advanced Manufacturing and Engineering Young Individual Research Grant (M4070267.120) by the Science \& Engineering Research Council of Agency for Science, Technology and Research Singapore, and Green and Sustainable Manufacturing Trust Fund 2013 by GlaxoSmithKline (Singapore). We are grateful to the National Supercomputing Centre (NSCC) of Singapore for supporting the numerical calculations.
\end{acknowledgments}

\bibliographystyle{h-physrev}
\bibliography{reference}

\clearpage

\clearpage

\end{document}